\documentclass{emulateapj}
\usepackage{amsmath,hyperref}

\shorttitle{Data Cubes and Design}
\shortauthors{Yeremi et al.}

\begin{document}


\title{Comparing Simulated Emission from Molecular Clouds Using Experimental Design}


\author{Miayan Yeremi\altaffilmark{1}, Mallory Flynn\altaffilmark{1}, Stella Offner\altaffilmark{2,3}, Jason Loeppky\altaffilmark{1}, Erik Rosolowsky\altaffilmark{1,4}}
\altaffiltext{1}{University of British Columbia, Okanagan
  Campus, Departments of Physics and Statistics, 3333 University
  Way, Kelowna BC V1V 1V7 Canada}
\altaffiltext{2}{Yale University Astronomy Department, 260 Whitney Ave, New Haven, CT 06511, USA}
\altaffiltext{3}{Hubble Fellow}
\altaffiltext{4}{University of Alberta, Department of Physics, 4-181 CCIS, Edmonton AB T6G 2E1, Canada}



\begin{abstract}
We propose a new approach to comparing simulated observations that enables us to determine the significance of the underlying physical effects.  We utilize the methodology of experimental design, a subfield of statistical analysis, to establish a framework for comparing simulated position-position-velocity data cubes to each other.  We propose three similarity metrics based on methods described in the literature: principal component analysis, the spectral correlation function, and the Cramer multi-variate two sample similarity statistic.  Using these metrics, we intercompare a suite of mock observational data of molecular clouds generated from magnetohydrodynamic simulations with varying physical conditions.  Using this framework, we show that all three metrics are sensitive to changing Mach number and temperature in the simulation sets, but cannot detect changes in magnetic field strength and initial velocity spectrum.  We highlight the shortcomings of one-factor-at-a-time designs commonly used in astrophysics and propose fractional factorial designs as a means to rigorously examine the effects of changing physical properties while minimizing the investment of computational resources.
\end{abstract}


\keywords{ISM:clouds --- methods: statistical --- radio lines:ISM}



\section{Introduction}
The process of star formation in molecular gas is often described as multi-physics.  A myriad of physical effects are thought to be important at regulating the process, through establishing the structure of molecular gas and the subsequent distribution of stellar masses that result.  Fluid dynamics shaped by gravitation, magnetic fields, chemistry, and radiation lies at the heart of the process and each of these effects is thought to contribute significantly to the outcome.  While analytic approaches guide our thinking, numerical simulations have been essential for modeling the nonlinear interaction of these effects.  As numerical simulations become more sophisticated, they provide a better emulation of the processes at work in molecular clouds.  By finding a good match between the observations and simulations, we recognize the underlying physical parameters governing the simulation as a possible (though not unique) set of conditions within molecular clouds.  

Given this improvement in numerical methods, many groups have begun to compare simulations of star forming molecular clouds to observations of the same \citep[e.g.,][]{falgarone-synth,padoan-scf,padoan-perseus,offner2008a,dib2010}.  These numerical studies all produced {\em synthetic observations}, which can be  compared to observations of actual molecular clouds.  The synthetic observation approach avoids the complexity of inferring the real physical properties of molecular gas from the incomplete information present in observations.  These comparisons usually leverage an underlying statistic to gauge the degree of similarity.  The most common route is to identify a population of objects within the simulation (e.g., dense molecular cores) and compare their properties to those of the known population.  The studies claim a successful emulation when the distributions of certain properties (mass, angular momentum, size, etc.) mimic those of the observed population.   Population comparisons work well when there is a well-defined population of observed objects to which the simulations can be compared. 

When comparing continuous structure such as that of the gas in a molecular cloud, different sets of statistics that rely on the continuous nature of the data can be employed.  In general, these statistics can be broadly divided into those that rely on the theoretical understanding of turbulence \citep{scalo-stats, kleiner-dickman,kleiner-dickman-meth,miesch-bally,vca,lp04, heyer-pca} and those inspired by observations \citep{houlahan1,falgarone-fractal,stutzki98, scf}.  In addition to those statistics that are constructed around the continuous nature of the star forming clouds, catalog-based approaches are commonly adopted.  The canonical approach is that of clump finding, which entails dividing or extracting a catalog of substructures from the molecular cloud (called clumps), and comparing the resulting population of objects \citep{gaussclumps, clumpfind,gaussclumps2}.  

With the improving quality of both simulations and observations, there is a growing need for a general framework by which the comparisons between observations and simulations can be rigorously assessed.  Many statistical tools are developed {\it ad hoc}, and it is unclear what each individual tool is telling us.  A given tool may not be sensitive to different physical parameters.  For example, fractal statistics may not be significantly altered by changes in the magnetic field.  Similarly, simulation efforts may claim to reproduce the structure of molecular clouds, but the assessment of that similarity is frequently qualitative. In this paper, we present a framework by which the sensitivity of different statistics to physical parameters can be measured.   We further use this framework to evaluate the quality of these statistics.   

We focus this work on the structure of star-forming molecular clouds.  Observations of molecular cloud structure use both spectral lines (typically CO) and the dust continuum.  Of these, we focus on the CO line mapping of clouds, which produces position-position-velocity (PPV) data cubes containing information about both the distribution and dynamics of molecular clouds.  Such PPV data sets display a wealth of structure, representing a challenging modeling problem for numerical simulations.  In this paper, we will outline a basic analysis framework and provide several candidate statistical measures for comparing data sets (\S \ref{section-comparison}).  We will then present a suite of numerical simulations that exhaustively probe a simple parameter space (\S \ref{section-simulations}).  Using this set of simulations, we describe how {\it linear models} provide a means to analyze the significance of the statistical measures (\S \ref{section-linearmodel}).  Using the results of the analysis, we critically evaluate the different proposed statistics in \S \ref{section:results}. 

This work emphasizes the comparison of a family of simulations to each other.  Future work will use this approach to compare simulations to observations and the optimization of simulation parameters to match observations.

\section{Comparison Statistics}
\label{section-comparison}

The comparison between simulations and observations is best carried out in the observational domain.  Our observational window onto reality is restricted to  information that can be inferred from the observed emission of these molecular clouds.  Inferring the full suite of physical cloud properties (density, temperature, velocities) from the emergent radiation is complicated by radiative transfer effects (particularly in optically thick emission), chemistry, spatially and temporally varying heating and cooling as well as imperfections in observations.  Compensating for these effects is difficult and always leaves caveats attached to the conclusions.  By creating synthetic observations from the known physics in simulations, we avoid having to invert these non-linear effects.  Instead, we rely on the same caveats affecting both our emulated and actual views of star formation.  Thus, a rigorous comparison of simulations and observations necessarily requires the comparison of synthetic observations with actual observations \citep[e.g.,][]{taste-testing}.  Even though the present work does not directly incorporate observational data, we operate on synthetic observations and deploy statistical tools that can also operate on observational data.  

Given two PPV data sets, $O_1$ and $O_2$, we propose a formalism based around a pseudo-distance metric that measures the differences between them.  Like other distance metrics, we require $d(O_1,O_2)\in \mathbb{R}^{0+}$ and, ideally, this pseudo-distance measures the difference in the underlying physics producing the data sets, such as temperature, magnetic field, and Mach number of the fluid flow.  In contrast, it should not be affected by other features such as coordinate shifts.  Indeed, a good similarity statistic will be unaffected by any representation effects such as the pixelization (though it should depend on the resolution).  For example, $d(O_1,O_2)=0$ if $O_2$ is just $O_1$ shifted along any of its axes, e.g., $O_2(x,y,v) = O_1(x+\delta x,y,v)$.  Any statistic that ignores the spatial relationships between pixels will trivially fulfill this criterion.  The distance metric should be sensitive to difference in physical scale for a scale factor $s$: $O_2(x,y,v) = O_1(sx,sy,v)$.  

For observational data, the statistic should be independent of the noise levels of the data set.  We denote the normal distribution with mean $\mu$ and variance $\sigma^2$ as $\mathcal{N}(\mu,\sigma^2)$.  For example, $d(O_1,O_2)\approx 0$ if $O_1 = O_{true}+\mathcal{N}(0,\sigma_1^2)$ and $O_2 = O_{true}+\mathcal{N}(0,\sigma_2^2)$ for the same $O_{true}$ but different $\sigma_1$ and $\sigma_2$. Practically, this means any statistic involving simulation data should not rely on any data that are at low intensity levels with respect to the maximum. Such data would be impossible to measure in real observational data sets.  The distance metric should be near zero for different realizations of the same physical processes.  In our notation, $d(O_1,O_2)\approx 0$ if $O_1$ and $O_2$ are synthetic observations generated with the same physical conditions but different random seeds.   Likewise, observations of regions in a similar evolutionary state should also show minimal difference under these statistics.  Finally, there is significant physical information in the spatial ordering of the pixels.  Self-gravitation, for example, leads to neighboring pixels begin correlated in their density and consequently their emission.  Thus, $d(O_1,O^\prime_1) \gg 0$ if $O^\prime_1$ is the same data as $O_1$ with the positions randomized (either in position or in position plus velocity).  Given these considerations, we propose several possible similarity statistics as follows.  Not all statistics fulfill all of the above criteria, and the different features they incorporate lead to different responses under the changing physical conditions (\S\ref{section:results}).

\subsection{Principal Component Analysis}
\citet{heyer-pca} first proposed an application of principal component analysis (PCA) to spectral line data cubes of molecular clouds.  Subsequently, PCA has been shown to be sensitive to the structure function of the underlying turbulent flow \citep{brunt-pca1,brunt-pca2}.  We adapt part of the PCA formalism to parameterize the physical differences between datasets.  Given two data sets, $O_1(x,y,v)$ and $O_2(x,y,v)$, we construct a covariance matrix, comparing the variance between different velocity channels in the data cube:
\begin{equation}
S_1(v,v^\prime) = \sum_{x,y} O_1(x,y,v) O_1(x,y,v^\prime),
\end{equation}
and a parallel construction for $O_2$ yielding $S_2$.  Following the PCA approach, we compute eigenvalues $\lambda$ and eigenvectors $u$ following $S u = \lambda u$.  To construct a distance measure, we sort the vectors eigenvalues in order of absolute value ($\boldsymbol{\lambda}_1$, $\boldsymbol{\lambda}_2$) and propose a distance metric:
\begin{equation}
d_{\mathrm{PCA}}(O_1,O_2) = \left[\frac{\sum_j(\boldsymbol{\lambda}_{1j} - \boldsymbol{\lambda}_{2j})^2}{\sum_j |\boldsymbol{\lambda}_{1j}| \sum_j|\boldsymbol{\lambda}_{2j}|}\right]^{1/2}.
\label{equation-dpca}
\end{equation}
Here, the sums are carried out over the elements of the vectors $\boldsymbol{\lambda}$.  The length of these vectors is equal to the number of velocity channels in the data.  In practice, nearly all of the power is found in the first $\sim 10$ eigenvalues, and thus we truncate the sums in Equation \ref{equation-dpca} after 50 terms.  This facilitates comparing the results to cubes with different extent in velocity.  
\subsection{The Spectral Correlation Function}
The Spectral Correlation function (SCF) was first proposed by \citet{scf}, and subsequently demonstrated to be a useful discriminant between the properties of turbulence in comparing simulations and observations \citep{padoan-scf}.  The SCF has a variety of different forms; here we express the SCF in terms of a normalized root-mean-square difference between spectra separated by an offset vector $\boldsymbol{\ell}$:
\begin{equation}
S(\boldsymbol{\ell}) = 1 - \left\langle \sqrt{\frac{\sum_v
  |O(\mathbf{r},v)-O(\mathbf{r}+\boldsymbol{\ell},v)|^2}{\sum_v
  |O(\mathbf{r},v)|^2+\sum_v |O(\mathbf{r}+\boldsymbol{\ell},v)|^2}}\right\rangle_{\mathbf{r}}.
\label{equation-scf}
\end{equation}
Here, $\mathbf{r}$ is a vector representing the two spatial coordinates within a datacube.  Note that $S(0)=1$ and generally decreases to zero as $|\boldsymbol{\ell}|$ grows large. Computing the SCF for a pair of data cubes will yield two different surfaces.  We calculate a distance metric as
\begin{equation}
d_{\mathrm{SCF}} = \sqrt{\sum_{\boldsymbol{\ell}} [S_1(\boldsymbol{\ell})-S_2(\boldsymbol{\ell})]^2}.
\end{equation}
Again, we truncate the summation over an area where the SCF is significant.  We chose a 23-pixel squared patch for calculating  $S$ and subsequently $d_{\mathrm{SCF}}$.

\subsection{The Cramer Statistic}
Motivated by the need for multivariate two-sample testing, the Cramer test statistic was first proposed by \citet{cramer-test}.
Given two dimensional data sets $P$ and $Q$ with sizes $N_P\times N_D$ and $N_Q\times N_D$ respectively, the test statistic is calculated by viewing each row as a point in a $N_D$-dimensional space.   The test statistic is computed by comparing the typical Euclidean inter-point distances between the data $P$ and the data in $Q$ to the distances between the data within each of $P$ and $Q$ respectively:  
\begin{eqnarray} d_{\mathrm{C}}(P,Q) & = &\frac{N_P N_Q}{N_P+N_Q} \left( \frac{1}{N_P N_Q}\sum_{p=1}^{N_P}\sum_{q=1}^{N_Q} || P_{pj} - Q_{qj} || \right. \nonumber \\ 
&&-\frac{1}{2N_P^{2}}\sum_{p_1,p_2=1}^{N_P} || P_{p_1 j}  - P_{p_2 j} || \nonumber \\
&&\left. -\frac{1}{2N_Q^{2}}\sum_{q_1,q_2=1}^{N_Q} || Q_{q_1 j} - Q_{q_2 j} || \right). \label{eq:cramer_met} \end{eqnarray} 
The above norm is the Euclidean norm between two points in $N_D$ dimensions: $||X_j - Y_j|| \equiv \sqrt{\sum_{j=1}^{N_D} (X_j-Y_j)}$. One advantage of the Cramer statistic is that its significance can be determined by bootstrapping.  The bootstrap draws two mock data sets of the same sizes as $P$ and $Q$ from the combined set $\{P,Q\}$ and evaluates the statistic over a large number of repetitions.  Additionally, the statistic has been shown to be more sensitive than the multivariate Kolmogorov-Smirnov test. 

In the context of the current work, we must recast PPV data cubes into sets of $N_D$-dimensional points suitable for the Cramer test.   We experimented with several methods for representing a PPV data cube as one or more $k$-dimensional points.  Given a data cube $O(x,y,v)$, one approach is to consider each spectrum with $N_D$ velocity channels as a datum and consider the sets $P$ and $Q$ as spectra drawn from the two different data cubes.  This comparison requires that the data sets being compared have matched spectral resolution and eliminates spatial information in the data.  Alternatively, we construct a vector from each velocity channel in the data cube by sorting the data into a vector of intensity values for each channel.  We found that only retaining the top quartile of the brightness data maintained the sensitivity of the statistic while improving performance.  We use the latter approach in this initial study and evaluate the application of the Cramer test to PPV data cubes in other work (Flynn et al., in prep.).  

The Cramer statistic is sensitive to the absolute values of the data set.  In order to eliminate this sensitivity, we normalize the transformed arrays by their respective spectral norms.   The spectral norm of an array $P$ is constructed by first taking the eigenvalues, $\boldsymbol{\lambda}$ of the square array $P^T P$.  The spectral norm is then $||P||_2 \equiv \sqrt{\max{\boldsymbol{\lambda}}}$.

We note that clearly the Cramer statistic is not sensitive to pixel ordering in our implementation.  The statistic performs well in the task of classifying similar and dissimilar data cubes, and experimentally satisfies the requirements of a well behaved distance metric. 

\section{Simulated Data Sets}
\label{section-simulations}

We ran 16 simulations of self-gravitating, magnetized turbulence using Enzo version 1.9 \citep{enzo}\footnote{John Wise kindly provided modifications to the sink particle code to prevent Enzo from crashing.}.  The simulations were designed to emulate a star forming cloud undergoing gravitational collapse.  We used a base grid with dimensions 128$^3$ over a box size of $L=10$ pc with periodic boundary conditions.  We set the initial density field of the box to be uniform $n_{\mathrm{H2}} = 36 \mbox{ cm}^{-3}$ and the initial direction of the magnetic field to be parallel to one face of the cube. For each of the simulations, we initially ran the simulations without self-gravity to allow the driving mechanism to develop turbulence.  The random driving pattern for the turbulence is solenoidal with power on wavenumbers from $k_v=3$ to $k_v=4$.  Depending on the run, we set the shape of the initial velocity spectrum to have power on a wide range of scale (from $k_v=2$ to $k_v=10$) or only large scale velocity structure ($k_v=1$ to $k_v=2$).   We scale the time of the simulation to the crossing time $t_c = L/\mathcal{M} c_s$ where $\mathcal{M}$ is the Mach number and $c_s$ is the sound speed.  At $t/t_c=2$, we turned on self-gravity, adaptive mesh refinement and enabled the formation of sink particles.  We used a Jeans length refinement criterion, refining the grid by a factor of two whenever the local Jeans length became shorter than four cells \citep{truelove97}.  We limited the number of refinement levels to four.  We systematically varied the input Mach number ($\mathcal{M}$, which sets the amplitude of the forcing field), initial magnetic field strength ($B_0$), initial velocity spectrum shape ($k_{v}$), and temperature ($T$) as described in Table \ref{table:codes}.  The table also reports the ``text codes'' we give to describe the varying parameters in the results.   We analyze the simulations at intervals of 0.1 crossing times from $t/t_c=2.3$ to $t/t_c=3.0$.

\begin{deluxetable}{ccccc}
\tablewidth{0pt}
\tablecaption{Coding for Parameters}
\tablehead{
\colhead{Level} & \colhead{$\mathcal{M}$} & \colhead{$B_0$} &\colhead{$k_{v}$} & \colhead{$T$}\\
\colhead{} & \colhead{} & \colhead{($\mu$G)} & & \colhead{(K)}
}
\startdata
Low & 6 & 1 & $[2,10]$ & 10 \\ 
High & 18 & 100 & $[1,2]$ & 30 \\
\tableline
\multicolumn{5}{c}{Text codes} \\
\tableline
& M & B & k & T
\enddata
\label{table:codes}
\end{deluxetable}

\begin{figure*}
\plotone{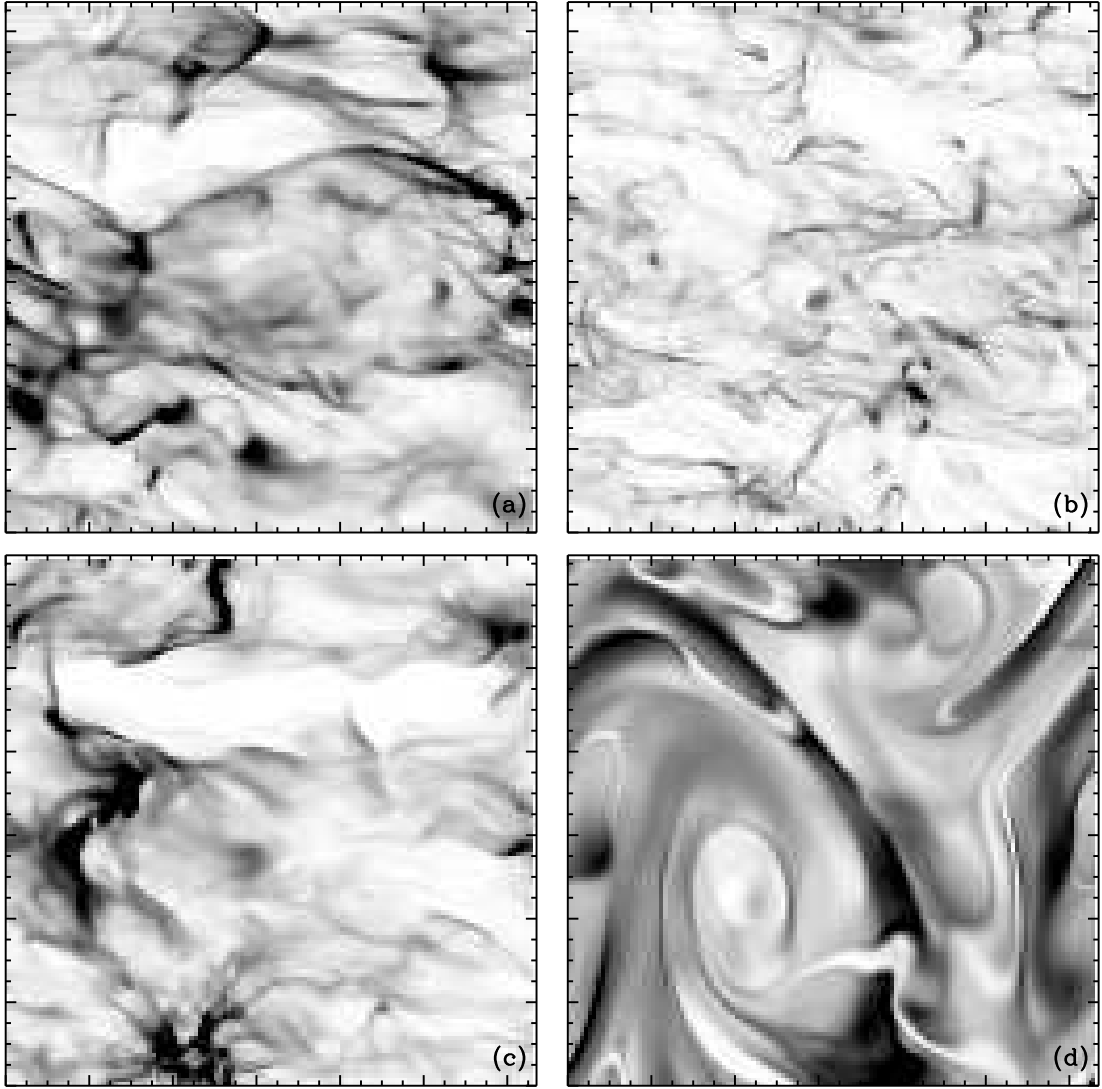}
\caption{\label{fig:compcube} Slices at $v=0$ through example PPV data cubes at $t/t_c=3.0$ in the study looking along the initial magnetic field direction.  The grayscale displays $^{13}\mbox{CO} (1\to 0)$ emission and runs linearly from 0 to 1.5 K.  Panel (a) shows the fiducial simulation with $\mathcal{M}=6$, $B_0=1~\mu\mathrm{G}$, $T=10$~K and a full range of initial velocity spectrum.  Panel (b) shows the effects for $\mathcal{M}=18$ and all other parameters as seen in (a).  Panel (c) shows the effects of changing the initial velocity spectrum only and panel (d) shows the effects of a strong magnetic field ($B_0=100~\mu\mathrm{G}$).}
\end{figure*}

We ran one simulation for each of the $2^4$ possible combinations of the parameter settings.  We post-processed the data using the RADMC-3D code \citep{radmc-3d} to produce PPV data cubes of $^{13}\mbox{CO} (J=1\to0)$ emission.  In RADMC-3D, the level populations are calculated using the large velocity gradient (LVG) approximation \citep{shetty-xfac}.  The $^{13}$CO collisional and emission coefficients are taken from the Leiden Atomic and Molecular Database \citep[LAMDA;][]{lamda}.  The simulated observations are generated as if the simulation domain were located a distance of 260 pc \citep[the distance of the Perseus molecular cloud, chosen for future comparison with observational data from the COMPLETE survey;][]{complete-data}.  We also selected $^{13}$CO as a tracer since it highlights the main structure of a molecular cloud while not being as limited by opacity effects in the $^{12}$CO data.   In the radiative transfer, we include ``microturbulence'' with $\sigma_v = 0.1\mbox{ km s}^{-1}$.  Microturbulence represents turbulent motions on scales too small to be resolved by the hydrodynamic grid and contributes to line broadening.  The velocity resolution of the data cubes is $\delta v=0.017\mbox{ km s}^{-1}$.  We produced PPV data sets for each of the three orthogonal orientations of the cube to capture the effects of observations along and parallel to the initial magnetic field.  Regions where the adaptive mesh undergoes refinement are averaged over so that the PPV data have the same spatial size as the base grid, namely 10 pc resolved into 128 resolution elements.  Our final data suite consists of (16 simulations)$\times$(8 time steps)$\times$(3 orientations)$=384$ synthetic PPV data cubes.  Figure \ref{fig:compcube} shows the $v=0$ slice through the data cubes for four different parameter settings at $t/t_c=3.0$.  Additional data from the simulations are presented in Appendix \ref{appendix-simprops}.  

\section{Response Analysis}
\label{section-linearmodel}

The work leverages the framework from the statistical subfield of experimental design.  The goal of this analysis is to determine how changes in physical conditions manifest in the synthetic data.   We refer to these changes as {\em effects} and aim to estimate their significance.  To make these estimates, we use similarity metrics to compare the values at the fiducial (or low) parameter values to those at the high values (Table \ref{table:codes}).  We define a scalar response when comparing the synthetic data from one simulation to another.  Since there are multiple timesteps available for each set of physical parameters, these need to be aggregated into a single statistic typifying how distinct simulations with different parameters are.   We define a scalar response value for the $i$th simulation compared to a fiducial simulation as
\begin{equation}
y_i = \frac{1}{N_t} \sum_{j=1}^{N_t} d[O_1(t_j),O_i(t_j)].
\label{eq:response}
\end{equation} 
This is the mean of a given comparison statistic over the last $N_t = 8$ outputs of the simulation (i.e., from $t/t_c=2.3$ to 3.0).  Each simulation is compared to the fiducial simulation $O_1$, which has $\mathcal{M}=6$, $B_0=1~\mu\mathrm{G}$, $T=10$~K and $k_v=[2,10]$.  The averaging occurs when self-gravity is active, and some simulations are evolving significantly.  The use of an average over time aggregates these effects, but the time evolution can be significant (see \S\ref{subsection-evolution}).

\subsection{Linear Models}

Typically, parameter studies of star formation use ``one factor at a time'' (OFAT) variation of parameters \citep[e.g.,][]{shetty-xfac, burkhart-bispectrum, atwood-velocity, vasquez-velocity-field,offner-h1h2}.  One parameter (e.g., the magnetic field strength) is varied while all the other factors in the simulation are held constant.  The response of the simulation to the variation of this single parameter is then interpreted as the effect of changing this parameter.  This approach is reasonable when simultaneous parameter changes will not result in a different degree response relative to individual changes, i.e., when there are no interactions between the parameters.  Analyzing a suite of OFAT parameter studies may be misleading when parameter interactions exist.

We establish quantitative measures of parameter changes by studying the resulting runs as a standard linear model.  We consider the case where we have $n$ different numerical simulations and $k$ different combinations of parameter settings.  Note that repeated experiments with the same parameter settings but different random seeds will increase $n$ but not $k$.  In matrix form we have 
\begin{equation}  \mathbf{Y} = \mathbf{X}\boldsymbol{\beta}+\boldsymbol{\varepsilon}, \label{eq:lin_mod} \end{equation} 
where $\mathbf{X}$ is an $n\times (k+1)$ model matrix, $\mathbf{Y}$ is the $n\times 1$ vector of responses (i.e., equation \ref{eq:response}), 
$\boldsymbol{\varepsilon}$ is the $n \times 1$ vector of independent normal errors and $
\boldsymbol{ \beta }^T=(\beta_0,\beta_1, \ldots, \beta_k)^T$ is the $(k+1)\times 1$ vector of unknown parameters (including the intercept).  These fit parameters characterize the response of the simulations to changing the physical conditions in the simulation.  In this framework, the OFAT design has a model matrix
\begin{equation} \mathbf{X} = \left( 
\begin{array}{ccccc}
 1 & 0 & 0 & 0 & 0  \\ 
 1 & 1 & 0 & 0 & 0  \\ 
 1 & 0 & 1 & 0 & 0  \\  
 1 & 0 & 0 & 1 & 0  \\  
 1 & 0 & 0 & 0 & 1   
\end{array} \right),
\label{eq:base_ofat}
\end{equation}
where the first column corresponds to a constant intercept, while the latter four columns code the factor setting for the four factors used in this study (see Table \ref{table:codes}). In OFAT, we represent the low/fiducial settings for parameters with a {\em code} of 0 and the high setting of parameters with a code of 1.  We utilize codes instead of the actual parameter values for simplicity in developing the design matrix (Equation \ref{eq:base_ofat}) and interpreting the statistical properties thereof.  There is substantial latitude in choice of coding since the values of $\boldsymbol{\beta}$.  Under certain coding schemes, the values of these effect estimates can be interpreted in terms of quantitative effects \citep{box05}, but here we are only interested in the significance of the values.  


The standard least squares estimator, $\hat{ \boldsymbol{\beta }}$, of the fit parameters $\boldsymbol{\beta}$ is given by: 
\begin{equation} \hat{ \boldsymbol{\beta }} = (\mathbf{X}^{T} \mathbf{X})^{-1} \mathbf{X}^{T}  \mathbf{Y}.   \label{eq:lse}    
\end{equation}

The covariance matrix for $ \boldsymbol{\beta} $ is given by
\begin{equation} \mbox{cov} ( \hat{ \boldsymbol{\beta }})  = \sigma^{2} ( \mathbf{X}^{T} \mathbf{X} )^{-1}, \label{eq:cov} \end{equation} 
where $\sigma^{2}$ is the unknown error variance characterizing $\boldsymbol{\varepsilon}$.  Determining $\sigma^2$ usually requires running multiple experiments with the same physical parameter settings but different random seeds.  The repeated experiments indicate what fraction of the measured variation is the result of statistical fluctuations, enabling these effects to be distinguished from changes that can be attributed to changing simulation parameters.
The covariance matrix for the parameters $\boldsymbol{\varepsilon}$ is given by 
\begin{equation} \sigma^2(\mathbf{X}^{T} \mathbf{X})^{-1} = \sigma^2\left( 
\begin{array}{rrrrrrr}
   1 & -1 & -1 & -1 & -1 \\ 
  -1 & 2 & 1 & 1 & 1 \\ 
  -1 & 1 & 2 & 1 & 1 \\ 
  -1 & 1 & 1 & 2 & 1 \\ 
  -1 & 1 & 1 & 1 & 2 
\end{array} \right)
\label{eq:cov_ofat}
\end{equation}
indicating that the parameter estimates in the OFAT approach are correlated with each other meaning that changes in one will affect changes in another parameter.  In the case of an experimental setup and measuring of parameter significance it is desirable to have independent (i.e., uncorrelated) parameter estimates.  Additionally, the above model does not allow for interactions between the parameters to be studied through that set of simulations.

\subsection{Full Factorial Designs}
We run a {\em full factorial} design where all possible changes of the parameters are considered.  Using only two values for each of the four factors, this setup results in $2^4=16$ runs.  In order to aid the resulting analysis, we adopt a different coding for the levels, which utilizes the linear independence between the columns of orthogonal arrays to produce independent estimates of the importance of different physical effects.  We note again that the choice of coding is arbitrary as we only seek to measure the significance of the effect estimates.  The fiducial/low will be coded as $-1$ (previously 0) and a $+1$ is used for the high level.  The first five columns of the model matrix are given as 
\begin{equation} \mathbf{X} = \left( 
\begin{array}{lr}
\begin{array}{rrrrr}
1& -1 & 1 & 1 & 1    \\ 
 1&1 & 1 & 1 & 1   \\ 
 1&-1 & -1 & 1 & 1  \\  
 1&1 & -1 & 1 & 1  \\
 1&-1 & 1 & -1 & 1  \\ 
 1&1 & 1 & -1 & 1   \\ 
 1&-1 & -1 & -1 & 1   \\ 
 1&1 & -1 & -1 & 1   \\ 
 1&-1 & 1 & 1 & -1  \\ 
 1&1 & 1 & 1 & -1   \\ 
 1&-1 & -1 & 1 & -1    \\ 
 1&1 & -1 & 1 & -1    \\ 
 1&-1 & 1 & -1 & -1   \\ 
 1&1 & 1 & -1 & -1   \\ 
 1&-1 & -1 & -1 & -1    \\   
 1&1 & -1 & -1 & -1   
\end{array} 
& \cdots \\
\end{array}
\right)
\label{eq:ortho_full}
\end{equation} 
where the first column represents the intercept and the remaining 4 columns show the level setting for the parameters.  For example the third row shows that Enzo would be run with $\mathcal{M}=6$, $B_0=1~\mu\mathrm{G}$, $k_v=[1,2]$ and $T=30$~K.  Also notice that rows 15, 16, 14, 10 and 7 respectively are the 5 runs in the OFAT design described in Equation \ref{eq:base_ofat}.

The above design allows for the estimation of all of the two, three and four factor interactions between the resulting parameter settings.  These terms are included in the model matrix by adding an additional 10 columns to $\mathbf{X}$ that are constructed by taking the product of all possible subsets of the columns 2 through 5 in Equation \ref{eq:ortho_full}.  When using this parameterization, $\mathbf{X}^{T} \mathbf{X} = n \mathbf{I}$ where $\mathbf{I}$ is the identity matrix, so we describe this coding as an {\em orthogonal design}.  In this case, the covariance matrix (corresponding to Equation \ref{eq:cov_ofat}) is diagonal, and the estimated effects $\hat{ \boldsymbol{\beta }}$ are statistically independent.  Under this framework each parameter estimate is given by $\hat \beta_i=\frac{1}{2}(\bar{\mathbf{y}}_{+} - \bar{\mathbf{y}}_{-})$ where $ \bar{\mathbf{y}}_{+}$ and $ \bar{\mathbf{y}}_{-}$ are the means of the responses at the high and low values of the parameter respectively (as defined in Table \ref{table:codes}).



The goal is to evaluate the statistical significance of these effects. Without a measure of the run-to-run variation in the response statistics, any derived effect could just be the result of random fluctuations.  Ideally, there would be several replicates of each experimental run which would allow for the estimation of $\sigma^2$ in the model.   Replicated simulations are runs with the same physical parameters, but different random seeds for the turbulent driving.  With an estimate of $\sigma^2$, the statistical significance of each of the regression coefficients ($\hat{\beta}_i$) can be assessed to determine what parameters are causing the real changes in the underlying PPV data cubes.   Usually, replicated numerical experiments entail a considerable investment of computational resources and using those resources to produce a different experimental run can provide substantially more information.

In the case of the two-level full factorial design, parameter significance can still be determined even without replicated runs.  \citet{lenth} introduced a formal approach to judge parameter significance in unreplicated experiments.  Lenth's method assumes that most effects and interactions are insignificant so they can be used to estimate the ``noise'' in the effect estimates.  This assumption is checked after analysis to verify that only a small number of significant effects are indeed significant.  The method cannot be relied upon if it finds nearly all or none the effects are significant.  A pseudo-standard error is calculated using a robust estimator $\Psi$  given by 
\begin{equation}
\Psi=1.5\cdot\text{median} \{  | \hat\beta_{i} | : {| \hat\beta_{i} |< 2.5 s_{0} } \} 
\label{eqn:pse}
\end{equation}
where $s_0=1.5\cdot \mathrm{median}\{ |\hat\beta_{i} |\}.$ The approach proceeds by constructing statistic similar to a Student's t: $\tau_i=\hat \beta_i/\Psi, i=1,\dots,k$ and comparing $\tau_i$ to a critical value set by the experimental design.  In the case of a 16 run experiment with 15 factors, any $|\tau_i|>2.16$ is judged as significant, where the critical value is determined using the so-called individual error rate and changes depending in the number of runs \citep{wu-hamada}.  

\section{Results}
\label{section:results}

Using the simulations described above, we calculate response vectors $\mathbf{Y}$ for each of the three similarity metrics described in \S\ref{section-comparison}.  In this section, we examine how the different similarity metrics respond to changes in the physical parameters defined in the initial conditions of the simulation.  We begin by examining what can be determined from the full suite of 16 simulations, representing all parameters and then we examine how the results change for an OFAT study.  Finally, we propose using fractional factorial designs, where a carefully chosen subset of the 16 simulations is used in the analysis.  

\subsection{Full Factorial Design}
\label{section-fullfac}
Using the full factorial design, we can estimate the effects $\hat{\beta}_{i}$ and their statistical significance.  The effects measure how responsive a statistic is to changing physical parameters in the simulations.  The influence of the four principal parameters (known as {\em main effects} in the design literature) and their interactions is probed with a full factorial design.  For example, the analysis measured how the SCF statistic responds to changes in input Mach number and temperature as well as the interaction term between these effects (coded as M:T) that measures the change in the response statistic in excess of the change seen from the main effects alone.  These additional interaction terms probe cases where the physical parameters amplify the effects of each other.  

\begin{figure*}
\plotone{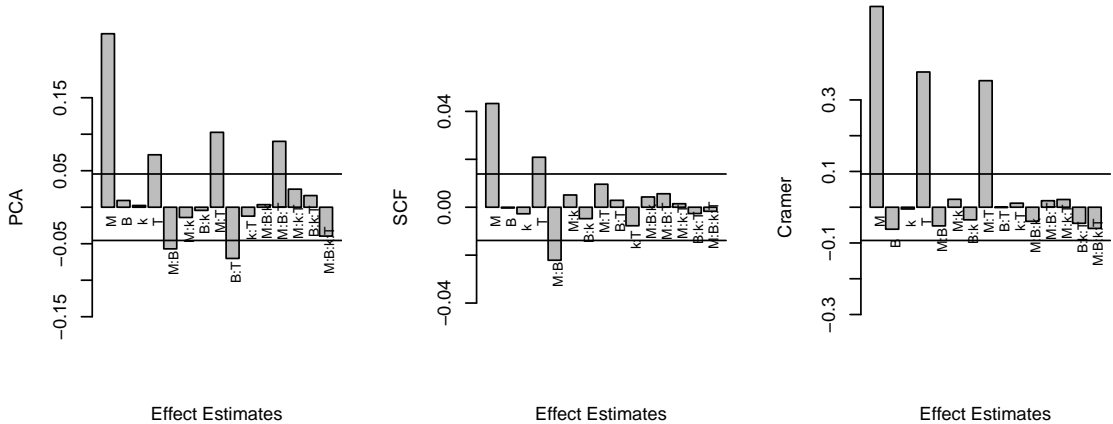}
\caption{ Lenth method estimates of parameters $\beta_i$ and their significance for full factorial design on the three test statistics: PCA, SCF, and Cramer.  Text codings for the different parameter effects are described in Table \ref{table:codes}.  The horizontal lines show the critical values of the Lenth statistics ($\pm 2.16\,\Psi$) and thus any effect bar that extends past the solid lines is considered significant.  Single letters (e.g., M, k) indicate main effects whereas combinations of letters (e.g., M:B or M:k:B) indicate interactions between the  effects.  All three diagnostics measure the input Mach number as the dominant effect across all the simulations, but the similarity statistics show differing sensitivities to other effects. \label{fig:full_lenth}}
\end{figure*}
 
Figure \ref{fig:full_lenth} shows the analysis of the full factorial design across all three similarity metrics.  The coefficients $\hat\beta_i$ are shown as the lengths of the bars.  The differences in the range of values shown for each $\hat{\beta}_i$ arises because the individual statistics have different numerical ranges in their response.  Only their value relative to the significance threshold of $2.16\Psi$ for each statistic is important.  The degree by which $\hat\beta_i$ exceeds the threshold can be interpreted as a degree of significance.  For example, the scaling of the three statistics with changing the input Mach number is more significant than any other effect for the three response statistics.  The analysis of the three comparison statistics consistently highlights two significant main effects: input Mach number and temperature.  Notably, there is no significant effect due to changing magnetic field alone.

Since the input Mach number of the turbulent driving in the simulation determines the density contrasts in shocks, this result is not surprising.   Note that the input Mach number is most closely tied to the amplitude of the turbulent forcing field and does not mean that actual Mach number of the flows in the simulation is this value (see \S\ref{subsection-driving}).  The temperature changes manifest in slightly larger line widths due to thermal pressure, but these effects should be insignificant because the sonic Mach numbers are large. The temperature increase also controls the amplitude of the spectral lines after radiative transfer (see Appendix A).  While the amplitude of the spectral lines is compensated for in the three statistics under comparison, the compensation has varying degrees of success.  The full-factorial analysis  indicates an interaction effect between the input Mach number and the temperature which is significant for PCA and Cramer statistics but not for the SCF.   The interaction effect is completely eliminated because of the nature of the SCF normalization which compensates for absolute values of the brightness in the cube comparison on a comparison-by-comparison basis.   The Cramer statistic also normalizes the data but appears to be more sensitive to variations due to the brightness of the data sets.  

Given the strong effects of the magnetic field visually apparent in Figure \ref{fig:compcube}, it is surprising that the magnetic field does not manifest as an active parameter in the response statistics.  This indicates a shortcoming of the statistics as formulated in the analysis: they are not sensitive to magnetic effects.  Even though the principal analysis is formulated for a line of sight parallel to the initial magnetic field, we performed a parallel analysis for the two faces perpendicular to the magnetic field and found no significant magnetic effects.  This lack of sensitivity to the strong effect on the emission structure in the highly magnetized case can be explained because the spectral line profiles for the $B_0=100~\mu\mathrm{G}$ are not significantly different from the fiducial case (see Figure \ref{fig-samplespec}).  Again, the analysis highlights the relative merits of the comparison statistics and indicates different statistics will need to be developed in order to measure the sensitivity to magnetic field.  Despite this lack of sensitivity to the main effect, both the PCA and SCF indicate a negative interaction effect between the magnetic field and the input Mach number.  This interaction indicates that joint increases in both the input Mach number and magnetic field produce simulations that these similarity measures find are more similar to the fiducial value than if these parameters are increased independently.  We attribute these significant interaction effects to a difference between the input Mach number, which is the parameter in the simulation, and the actual Mach number which establishes the density structure in the simulation (see \S\ref{subsection-driving}).  

Despite the comparatively low resolution of the simulation suite, the analysis of the full factorial design indicates the initial success at adopting the formalism.  The results show that the framework can identify some of the significant physical effects (input Mach number and temperature) that govern the structure of PPV data cubes.  However, the nondetection of effects that are clearly significant (magnetic field) shows that the construction of the similarity statistics needs to be adapted to detect magnetism.  Alternatively, other similarity measures which emphasize magnetic effects could be recast into this framework \citep[e.g.,][]{anisotropy-pca,burkhart-dendrograms}.  We emphasize that the results identify the significance of both the effects and their interactions, representing the primary utility in employing the fully developed statistical framework to this analysis.

\subsection{One Factor at a Time}

\begin{table} 
\begin{center}
\begin{tabular}{rrrr}
 \hline
Code & PCA &SCF & Cramer \\ 
 \hline
(Intercept) & 0.00 & 0.00 & 0.00 \\
 M& 0.80 & 0.14 & 0.57 \\   
 B & 0.26 & 0.10 & 0.06 \\ 
 k & 0.16 & 0.02  & 0.01 \\   
 T & 0.49 & 0.04 & 0.12 \\  
  \hline
\end{tabular}
\caption{Effect estimates of the various statistics based on the one factor at a time experiment (OFAT).}
\label{tb:ofat}
\end{center}
\end{table}

Where the full factorial analysis identifies significant main effects as well as their interaction terms, OFAT analysis can only detect main effects.  Furthermore, OFAT analysis without multiple realizations of the physical effects yields no estimate of the variance that arises because of the random fluctuations between simulations with the same physical conditions.  The OFAT analysis can be used to compare the magnitude of the main effects as given in Table \ref{tb:ofat} for the three different similarity statistics.  As with the full factorial design, input Mach number remains the dominant effect.  Temperature effects and magnetic effects also seem to be detectable and the initial velocity spectrum is least important affect.  The intercept value is estimated as 0.0 for all statistics because of the definition of the response, namely $y_1=0$ when the fiducial simulation is compared to itself.  We note that, in practice, the fiducial data will likely be observational data sets to which the simulations are being compared.  For analyzing a suite of simulations without observational data, a simulation at the center of the parameter space would make the best choice for a fiducial simulation.  

The effect estimate for the magnetic field seems like it could be detected by the similarity statistics; however, the full-factorial design indicates that this apparent detection arises because of the interaction effects seen between magnetic field strength and input Mach number.  A purely OFAT analysis will incorrectly conclude that the influence of the magnetic field strength has been detected.  Furthermore, OFAT provides no assessment of the variance in the similarity statistics and the importance of each of these effect estimates cannot be determined.  This analysis highlights the shortcomings of the OFAT approach.  Specifically, the analysis can be misled in the presence of parameter interactions and is inefficient at testing the significance of different simulation parameters \citep{box05}.  


\subsection{Fractional Factorial Design}

\begin{figure*}
\plotone{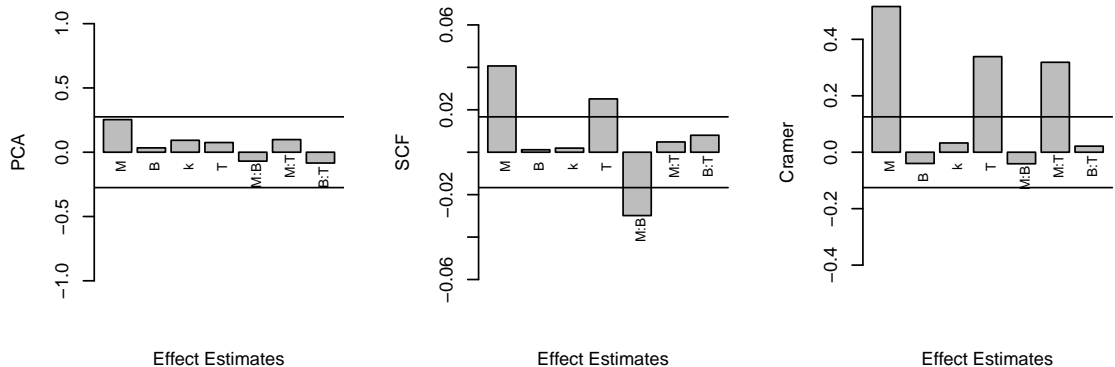}
\caption{ Lenth method for fractional factorial orthogonal design on the three test statistics, PCA, SCF, and Cramer.  Any parameter effect bar that extends past the solid line indicating the critical value for the statistics is considered significant.  The general behavior of the analysis agrees favorably with a full factorial design, even though the results are less significant.}
\vspace{0.2in}
\label{fig:frac_lenth}
\end{figure*}

The full factorial design provides a more nuanced interpretation of the responses as a function of changing initial conditions.  However, running a full factorial design can require substantial computational resources.   We ran the two-level, full factorial design, because we were using 128$^3$ resolution simulations. In addition, we chose a model where there are only four factors and each factor only has two values.  There are situations where one would like to run simulations with higher resolution, more factors, and a continuous range of values.  To reduce the computational cost, it is possible to run a subset of the full factorial design with the ability to estimate both the main effects and some of the interactions.   In many situations, prior knowledge can be used to identify a limited subset of the interactions that are likely to be important.  Starting from the full factorial design of 16 runs, we select the 8 runs that correspond to having all factors (M,B,k,T) at the high level.  This results in the following 8 runs, which are commonly referred to as a $2^{4-1}$ fractional factorial design.  
  \begin{equation} \mathbf{X} = \left( 
\begin{array}{rrrrrrr} 
1 & -1 & -1 & -1 & -1 \\ 
1 & 1 & -1 & -1 & 1 \\ 
 1 & -1 & 1 & -1 & 1 \\ 
1 & 1 & 1 & -1 & -1 \\ 
1 & -1 & -1 & 1 & 1 \\ 
1 & 1 & -1 & 1 & -1 \\ 
1 & -1 & 1 & 1 & -1 \\ 
1 & 1 & 1 & 1 & 1 \\ 
\end{array} \right).
\label{eq:ortho_ffd}
\end{equation} 
Again the columns 2 through 5 are used to assign the level setting for each of the four factors.  Since we have reduced the run size by a factor of two we lose the ability to estimate 8 of the 15 effects that were estimated in \S\ref{section-fullfac}.  Of the 7 remaining degrees of freedom, one can attribute the significance of the estimated effect to one of two terms.  For example, if the effect of column 2 is significant then either the main effect of input Mach number (M) or the B:k:T interaction is important.  We assume third and higher order terms are negligible, so we can estimate each of the four main effects.  The remaining three columns can be used to estimate the following pairs of effects, M:B=k:T, M:T=k:B and B:T=k:M.  The notation M:B=k:T is an indication that, in fractional factorial design, estimates of the two-factor interaction M:B are aliased with the k:T interaction.  If significant, we would not be able to discriminate between these two effects.  This aliasing is a cost of running a factional factorial design.  However, we can use our prior knowledge to decide whether a given significant effect should be interpreted as, for example, a M:T interaction or a k:B interaction.   We believe that the effects from the initial velocity spectrum are negligible because those effects are washed out by the turbulent driving.  Our assumption is well justified by the insignificance of velocity spectrum effects in the full factorial response analysis.  In larger simulations, it is possible the velocity spectrum and magnetic field could show dependence.  Thus, for this set of simulations, we believe that the significant interactions would be attributed to M:B, M:T and B:T as opposed to the effects involving the initial velocity spectrum (k).  Ignoring the aliasing of the parameters in the above design, the 8 runs simply correspond to a full factorial in three factors. Hence, one can use the same unreplicated analysis proposed by \citet{lenth} to determine which parameters are significant.  In this case the appropriate critical value is $|\tau|=2.279$.  Figure \ref{fig:frac_lenth} shows the results from the fractional factorial analysis. 
  
The significant parameters from both the full factorial and fractional factorial are consistent in relative magnitude and sign.  The effect significance, particularly for the PCA is reduced under the fractional factorial design.  Some information is lost when the fractional factorial is used over the full factorial,  but the fractional factorial does provide some information about the interactions.  In addition, it only requires an additional 3 runs in comparison to the OFAT design, while establishing an understanding of the parameter interactions that compares well to running the full suite simulations at twice the computational cost.  The results from this design illustrate how a carefully chosen design can maximize information from a parameter studies that are subject to limited computational resources.   The savings of a fractional factorial design over a full factorial grows as the number of parameters explored or the number of level settings increases.  

\section{Discussion}
\label{section-discussion}
\subsection{Driving and Mach Number}
\label{subsection-driving}
The dominant physical effect we see in our simulations is the response of our statistics to the input Mach number parameter set.  This input Mach number does not actually set the true Mach number of the fluid flow in the simulation.  Rather, it determines the amplitude of the driving field used to establish the turbulent velocity field \citep[][outlines an approach similar to that used in Enzo]{maclow99}.  The default Enzo distribution uses a heuristic formulation to specify the amplitude of the turbulent driving force per unit mass: $f = \eta \mathcal{M} c_s v_{rms}^{-1} t_{ff}^{-1}$, where $\eta$ is a {\em driving efficiency}, $c_s$ is the sound speed, $v_{rms}$ is the initial rms velocity measured in the simulation and $t_{ff}$ is the mean free-fall time in the domain.  The parameter $\mathcal{M}$ is the input Mach number for the simulation but does not necessarily mean that the Mach number of the simulation will have that value.  The driving efficiency was taken as $\eta=1600$ for all simulations, but this value only produces real Mach numbers close to the initial Mach number parameter for weak $B$-field cases.  The initial Mach number should thus be thought as a characteristic of the strength of the driving field rather than the true Mach number characterizing the simulation.  When compared to the input Mach number, we found that the actual Mach numbers were between 37\% ($B_0=100~\mu\mathrm{G}$, $k_v = [1,2]$) and 95\% (fiducial) of the input Mach number.  Of note, the input Mach number parameter also establishes the crossing time for the simulation.  The actual crossing time will be different from the crossing time reported in the simulation.  We also constructed a linear model of the actual Mach number in the simulation as a function of the changing input parameters.  As expected, the input Mach number was the dominant effect in establishing the real Mach number, but the initial magnetic field strength has a strong negative effect (increasing $B_0$ decreased actual Mach number relative to the input Mach number).  The damping effect of the magnetic field is expected on physical grounds and serves to explain why the magnetic field appears significant in OFAT analysis, but the effect is not present in the full-factorial design.   Instead, interaction effects between the magnetic field and temperature or input Mach number appear significant.   The interaction effects in the actual simulation response can be attributed partly to the magnetic field driving a discrepancy between input and actual Mach number.  Additionally, a modest Mach-temperature interaction emerged from the model which may explain the same effect seen in the response of the comparison metrics.  

\subsection{Time Evolution}
\label{subsection-evolution}

\begin{figure*}
\plotone{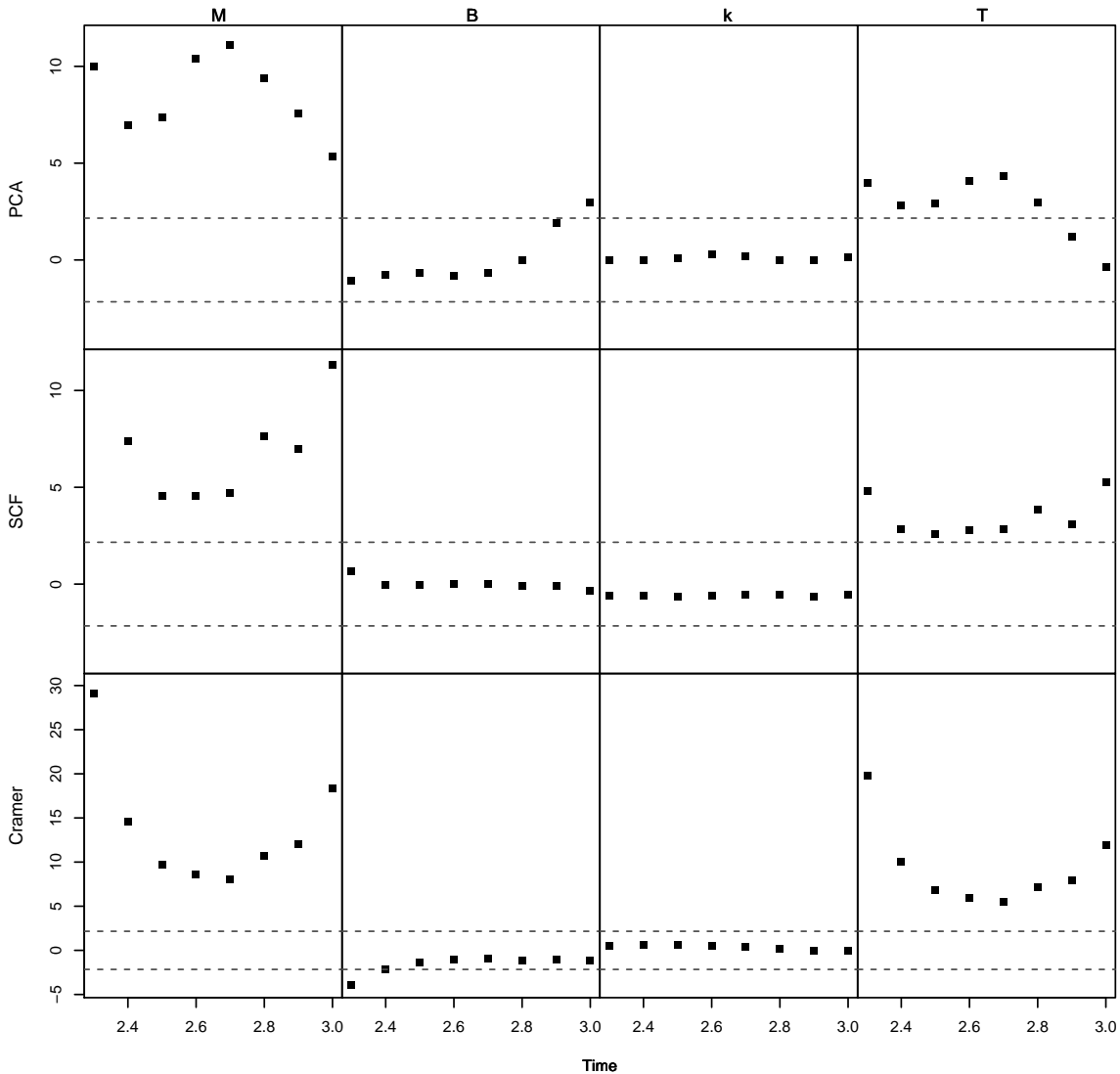}
\caption{\label{fig:timevar} Time variation of the main effect estimates for different timesteps in the simulation.  Individual points show the magnitude of the effect for changing each of the physical parameters -- Mach number (M), magnetic field strength (B), shape of the initial velocity spectrum (k), and temperature (T) -- on each of the statistics (Cramer, SCF and PCA) as a function of time in units of the crossing time $t_c$.  The dashed lines indicate the magnitude required for each effect to be deemed significant. }
\end{figure*}

Our primary analysis averages over the last eight time steps of the simulation (from $t/t_c=2.3$ to 3.0).  Since self-gravity is turned on at time $t/t_c=2.0$, simulations that are undergoing gravitational collapse will see these effects manifest over the time period contributing to the average.  Real effects may be averaged over or diminished by using the aggregate of all the response values over the self-gravitating phase of the simulation.  To address this concern, we plot the main effect estimates as a function of time in Figure \ref{fig:timevar}.  This figure compares each simulation at a fixed time step (though note that $t_c$ will vary by simulation based on the input Mach number as above).  Figure \ref{fig:timevar} shows that there is not a significant change in the main effect estimates over the period when the simulation is self-gravitating.  There is some variation in the sensitivity of the SCF to the magnetic field at the last timesteps of the simulation, but this effect is modest.  The significance of the effects for Mach and temperature changes appears significant over the whole period.

\section{Conclusions}

In this work, we have proposed a formalism by which PPV data cubes of spectral line data can be compared in order to assess physical effects.  The formalism relies on the methods utilized in the statistical subfield of experimental design.  We have proposed three similarity metrics based on existing techniques in the literature and assess their reliability using the formalism applied to a suite of simulations.  The three similarity metrics are principal component analysis, the spectral correlation function and a new method using the Cramer multivariate two-sample test.  All three similarity statistics showed consistent results in our analysis, though the relative sensitivities of these measures varied.  In particular, the statistics are most sensitive to Mach number and temperature effects.

We have also examined the benefits and liabilities for using the ``one factor at a time'' (OFAT) analysis that is common in the astrophysical literature.  While the results of OFAT  analysis are relatively easy to interpret, the analysis is unable to assess whether the physical effects explored in the simulations interact with each other.  This can lead to overinterpreting the importance of certain parameters in the simulation.  For example, we found that our measurement statistics appeared sensitive to magnetic field strength based on the OFAT analysis.  However, in exploring the parameter space fully, the statistics were found to be unable to detect changes in the magnetic field strength, instead finding an interaction effect between magnetic field strength and the input Mach number.  We attribute this interaction to the reduction of the real Mach number relative to the input for increasing magnetic field strength.  A full exploration of parameter space is invariably computationally prohibitive for high quality simulations.  Thus, we advocate selecting a fractional factorial design that is able to recover those effects most likely to be significant through the study.

A major shortcoming of this study is the exaggerated ranges used for the parameters (e.g., 100 $\mu$G fields on 10 pc scales are not observed).  Even so, this work demonstrates proof-of-concept for three avenues of study.  First, the comparison of simulations with observations can be explored under this framework.  Frequently, simulations are compared to observations and the conditions for a best match are assessed under qualitative terms.  By establishing observations as the fiducial to which the simulation suite is compared, we can establish those effects which most closely govern similarity to simulations.  While this would formalize the comparison between simulations and observations, it will require amending the similarity statistics to account for the effects of telescope observations (variable data set sizes, limited resolution, receiver noise). 

The second set of work that this study enables is the ability to assess the reliability of comparison metrics.  There are a wealth of such metrics present in the literature, but they are only validated under a limited set of conditions.  Using this test suite of simulations or a higher quality suite that studies a full range of parameter combinations, we can assess how reliable the simulations are at detecting certain physical effects.  The framework also reveals when a statistic can be conflating the effects of two parameters.  For example, the Cramer statistic is primarily sensitive to Mach and temperature effects and is surprisingly immune to interaction effects between these parameters and the magnetic field.  The PCA and SCF metrics detect significant statistical interactions between these parameters and the magnetic field strength.  However, none of these statistics detect the magnetic field as a main effect indicating limitations in their current formulation.

Finally, this study offers guidance for explorations of parameter space in suites of simulations.  Given the prevalence of OFAT analysis in astronomy and the complexity of interpreting the simultaneous variation of multiple physical effects, using a statistical response model to assess parameter effects offers some clear guidance.  Extensions to this framework allow for multiple values or continuous exploration of parameter spaces, non-linear response models and efficient exploration of high-dimensional parameter spaces for simulation. 

\acknowledgements This project was supported by Discovery Grants from NSERC of Canada and a Collaborative Research Grant from the I.K. Barber School of Arts and Sciences at UBC.  SSRO is supported by a NASA through a Hubble Fellowship grant \#51311.01 awarded by the Space Telescope Science Institute, which is operated by the Association of Universities for Research in Astronomy, Inc., for NASA, under contract NAS 5-26555.  We are grateful for simulation assistance from Tyler Berg.  We acknowledge productive discussions with Doug Johnstone and Alyssa Goodman in formulating this research.   WestGrid and Compute Canada resources were used in executing the simulations.  This research made heavy use freely-available, community-developed software, including Enzo (\url{http://enzo-project.org}),  yt \citep[\url{http://yt-project.org/};][]{yt-paper} and astropy \citep[\url{http://astropy.org};][]{astropy} and we are grateful for the developer efforts on those projects.  This research has made use of NASA's Astrophysics Data System.  We acknowledge the constructive comments from the anonymous referee, which improved the quality and presentation of the paper.  


\appendix

\section{Simulation Properties} 
\label{appendix-simprops}
\begin{figure}
\vspace{3in}
\plottwo{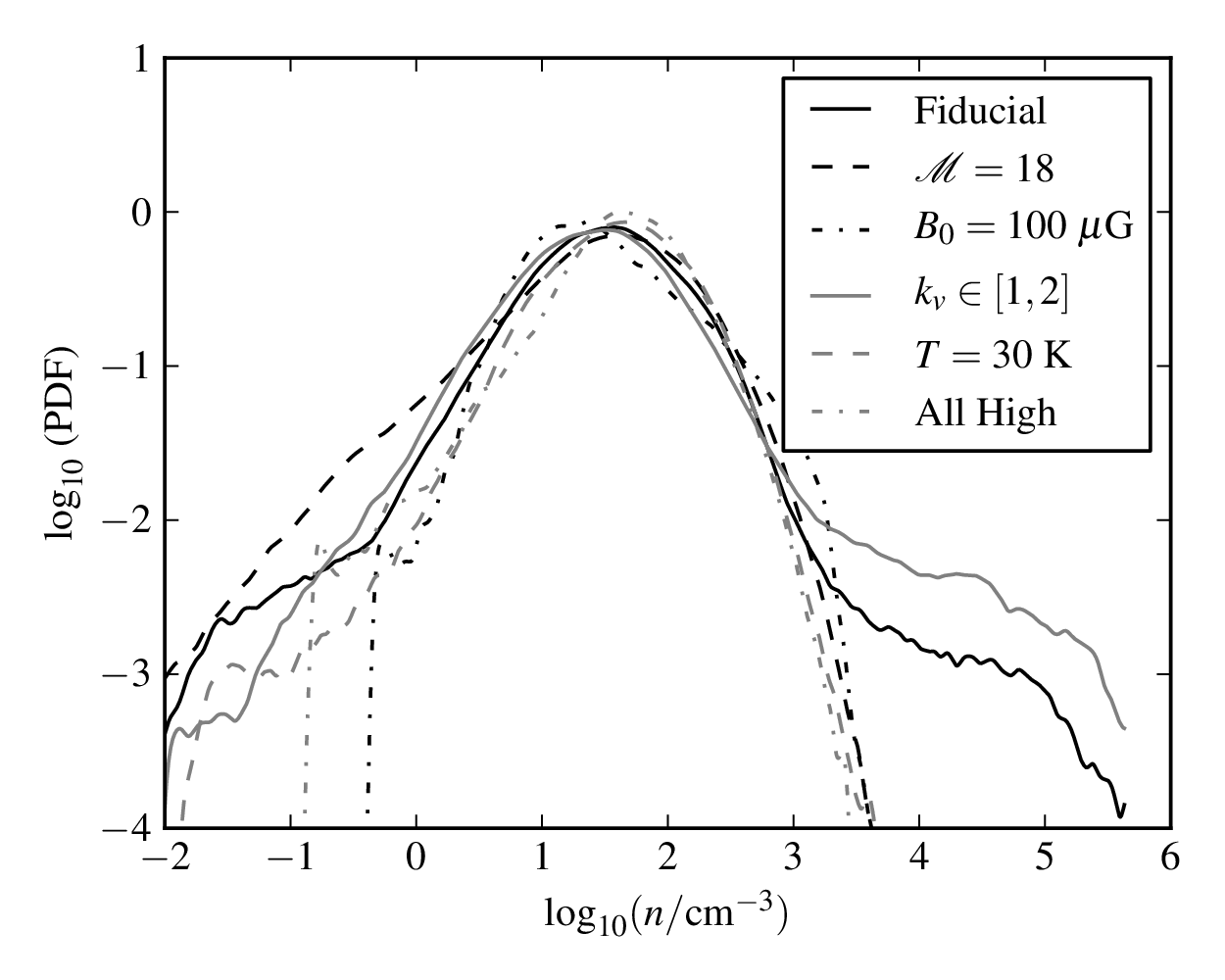}{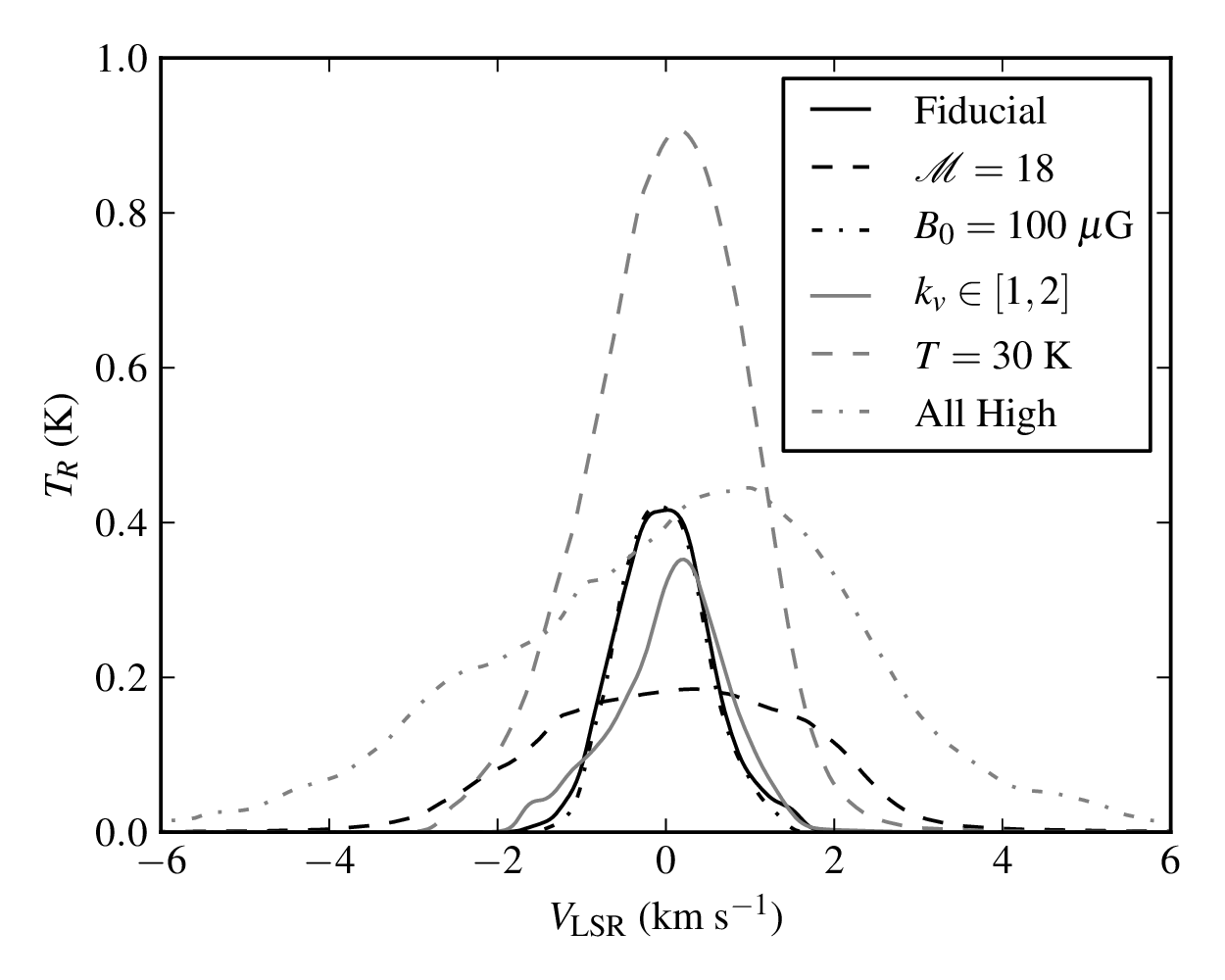}
\caption{\label{fig-samplespec} (left) Distribution of number density in the simulation volume and (right) average spectra from different simulations in the study.  The different curves show the properties for six of the 16 simulations.   The fiducial simulation has $\mathcal{M}=6$, $B_0=1~\mu\mathrm{G}$, $k_v = [2,10]$, and $T=10~\mathrm{K}$.  The other curves represent changing one of the simulation parameters to the value shown in the caption, except the last curve which illustrates setting all parameters to the high values.}   
\end{figure}

In Figure \ref{fig-samplespec} we present two summaries of the simulations in the simulation suite.  Rather than plotting curves for all 16 of the simulations, we only illustrate the fiducial case with all parameters at the low setting, the curves for only one factor changing from its low to high setting and finally the curves for all parameters set to their high value.  The two summaries present density PDFs and spectra for these simulations.  Of note, the density PDFs show that only the fiducial case and the large scale initial spectrum ($k_v= [1,2]$) show significant gravitational collapse and a tail of values off to high densities \citep{mlk04}.  In the other simulations, including those not shown in the figure, the local Jeans refinement condition is never met and there is no local gravitational collapse.  The initial driving scale likely plays little role at governing gravitational collapse since the fixed driving pattern dominates the structure of the turbulence at the time of gravitational collapse. 

The average spectra of the simulated observations also show roughly expected behavior.  The line width is larger for higher Mach numbers and temperature.  The amplitude of the line is also larger for the higher temperatures as expected from radiative transfer effects.  Spectra along individual lines of sight show significant non-Gaussian structure as is frequently seen in simulated data.  

\end{document}